\documentclass[aps,prd,amsmath,amssymb,superscriptaddress,nofootinbib,twocolumn,10pt,showpacs]{revtex4}
\usepackage[]{latexsym}
\usepackage[english]{babel}
\usepackage{graphicx}
\usepackage{hyperref}
\usepackage{color}
\usepackage{bm}
\usepackage{xcolor}

\allowdisplaybreaks

\begin{document}

\title{Interacting scalar fields: Dark matter and early dark energy}

\author{Gabriela Garcia-Arroyo}\email{arroyo@icf.unam.mx}\affiliation{Instituto de Ciencias Físicas,
Universidad Nacional Autónoma de México,
62210, Cuernavaca, Morelos, México.}

\author{L. Arturo Ureña-López}\email{lurena@ugto.mx}\affiliation{Departamento de Física, DCI, Campus León, 
Universidad de Guanajuato, 37150, León, Guanajuato, México.}

\author{J. Alberto V\'azquez}\email{javazquez@icf.unam.mx}\affiliation{Instituto de Ciencias Físicas,
Universidad Nacional Autónoma de México,
62210, Cuernavaca, Morelos, México.}


\begin{abstract}
The main aim of this work is to explore the possibility that cold dark matter (CDM) and early dark energy (EDE) can be described by canonical scalar fields that are coupled at the level of its conservation equations. The formalism covers dynamical aspects at the background and linear perturbation levels for an arbitrary coupling function, followed by an example of it. 
We emphasize the impact of this model on the Matter Power Spectrum and the Cosmic Microwave Background (CMB) spectra, with or without direct interaction. Our findings indicate that the presence of a scalar field can partially counteract the known effects of the other, opening the possibility to avoid some undesired aspects, such as the increase in $\Omega_{m}$ that usually is needed in the case of a purely EDE scalar field scenario, in order to fit the CMB spectra. This opens up the possibility to analyzing whether the interaction can help to ameliorate the cosmological tensions. \\

\end{abstract}

\maketitle

\begin{section}{Introduction}

One of the main conundrums of modern cosmology is the explanation of the dynamics of the Universe, 
led by two still-unknown components called dark energy and dark matter.
These ingredients are the foundations of the standard cosmological model, or $\Lambda$ CDM.
Here, the cosmological constant ($\Lambda$) plays the key role of dark energy and it is
considered to be responsible for the current accelerating expansion of the Universe,
suggested by many cosmological observations \cite{Perlmutter:1998np, Astier:2005qq};
whereas the cold dark matter (CDM) is the principal component for structure formation which 
significantly contributes to the observed rotation curves of galaxies \cite{Bernal:2017oih}.
Despite the simplicity of the model and that it provides a very accurate description to the 
majority of the astronomical observations, it carries out several issues of fundamental nature, i.e. at
the largest scales the cosmological constant problems \cite{Peebles_2003, Padmanabhan_2003} or the recent 
$H_0$ tension 
\cite{Abdalla:2022yfr}; at the galactic levels, the unexplained central density behaviors in halos and the 
overpopulation of small substructures \cite{Weinberg:2013aya, Bull:2015stt}. 
Therefore, a viable alternative to replace the standard
description of such components is the scalar fields, to behave as dark matter~\cite{Lee:1995af, Matos:1998vk, 
Matos:1999et, Hu:2000ke}, dark energy~\cite{Peebles:1987ek, Caldwell:1997ii,Steinhardt:2003st, Copeland:1997et} 
or inflation \cite{Guth:1980zm, Linde:1981mu, Albrecht:1982wi,Alcaniz:2006nu}.
However, to achieve these characteristics, it is still necessary to specify the potential of the 
field and its initial conditions.
\\

The idea of replacing cold dark matter for scalar fields was considered a couple 
of decades ago, with the simplest possibility of a real field, minimally coupled to gravity, and 
interacting only gravitationally with ordinary matter \cite{Ji1994,Lee1996,Guzman1997}.
The requirement for the scalar field, $\phi$, to mimic a pressureless fluid is to include 
a convex potential with a minimum at some value of the field, $\phi_c$, such that the mass of 
the associated particle could be nonnull. 
An example of such a type is the parabolic function $V(\phi)=1/2m_\phi^2\phi^2$ 
whose mass is defined as standard, $m_{\phi}^2=\partial_{\phi}^2 V(\phi_c)$ 
\cite{Matos2009, Suarez:2013iw}.
However, depending on the particular form of the potential, it may experience different 
behaviors before acting like a pressureless fluid, hence several alternatives are also brought into consideration, i.e., the self-interacting potential with a quartic term contribution 
$V(\phi)=1/2m_\phi^2\phi^2 + \lambda_\phi \phi^4$ \cite{Khlopov:1985fch, Padilla:2019fju, Li:2013nal, Suarez:2015fga}, 
the axionlike or trigonometric potential 
$V(\phi)=m_\phi^2f^2[1+\cos(\phi/f)]$ \cite{Ross:2016hyb, Cedeno:2017sou} and its analog 
$V(\phi)=m_\phi^2f^2[\cosh(\phi/f)-1]$ \cite{Matos2009DynamicsOS, Urena-Lopez:2019xri}.
The dynamics of these and some other potentials has been studied extensively in detail \cite{LinaresCedeno:2021sws}, as well as the combination of two scalar fields with 
different contributions to the total DM \cite{Tellez-Tovar:2021mge}. 
\\

With a similar idea in mind, scalar fields may also play an important role leading to 
dynamical dark energy models.
Here, the task is to be able to imitate the cosmological constant behavior at late times,
by including a minimally couple to gravity field 
with a kinetic energy term, which its positive sign corresponds to quintessence and the 
negative to phantom; and also a given potential.
The quintessence model is considered as the simplest scenario with no theoretical problems, 
such as the appearance of ghosts or Laplacian instabilities, that describes a dark energy evolving over time~\cite{Tsujikawa:2013fta, Linder:2007wa}.
These types of models are often classified into two broad categories such as ``thawing'' or
``freezing''~\cite{Caldwell:2005tm, Pantazis:2016nky} depending on their behavior over time.
A step further is to test a collection of potentials and compare its statistical viability
in terms of current observations \cite{PhysRevD.97.043524,PhysRevD.98.063530,Vazquez:2020ani, Banerjee:2020xcn}, 
or to include more than a single scalar field \cite{Cai:2009zp,Bamba:2012cp, Vazquez:2023kyx, Chimento:2008ws, vandeBruck:2022xbk}
to be able to reproduce the crossing of the phantom-divide line (PDL) shown by several model-independent reconstructions \cite{Escamilla:2021uoj,Vazquez:2012cen, Hee:2016nho, Zhao:2017cud}. 

Moreover, there is another type of scalar field potentials, named Early Dark Energy 
(EDE)~\cite{Doran_2006, Agrawal:2019lmo}, that could { have 
a 
non-negligible contribution during the early universe, prior to the onset of the current 
dark-energy-dominated epoch, where it is responsible of the accelerated expansion of the universe}. If the fraction of this early dark energy is large enough, it could 
have strongly affected the physics of the early universe and left its signature in the 
cosmic microwave background radiation and matter power spectrum. There are different EDE 
potentials, one of the most studied being the axion-like potential, denoted by
$V(\phi)=m^2f^2[1-\cos(\phi/f)]^n$. This potential has been inspired by ultralight axions (ULA) 
from string theory~\cite{Poulin:2018cxd, Doran:2006kp, Poulin:2023lkg}. It is important to 
note that ultralight axions are also candidates for dark matter~\cite{Hui:2016ltb, Rogers:2023ezo}.
Essentially, a ULA behaves as DE when its mass value is within the range
$m_{\phi}\lesssim 10^{-27}\rm{eV}$, whereas in the opposite mass range it dilutes away as DM.
In this work, without loss of generality, we focus on the EDE models with an 
Albrecht-Skordis (AS) potential; in~\cite{Barrow:2000nc, Albrecht:1999rm} the authors discuss some 
of its properties with emphasis in the cosmological eras and in view of the ability 
to explain the available observations of that epoch. A more recent work~\cite{Adil:2022hkj} also explores 
the ability of the model to solve the $H_0$ tension and the possible connections with different 
extensions of $\Lambda$CDM.
The dynamics of the AS potential allows us to have an almost constant EDE contribution during the early universe,
modulated by the potential's parameters. In fact, it mimics the behavior of radiation, then it 
begins to decay at a scale factor around $10^{-6}$, reaches a minimum at approximately $10^{-1}$ 
before transitioning into a growth phase to attain the characteristics of the late dark energy. 
Importantly, the AS potential does not require the inclusion of a cosmological constant.
\\

Even though single scalar field models provide a very good description of the evolution of 
the cosmological densities and peaks of the cosmic microwave background (CMB), as well as the number 
of substructures in galaxy arrays \cite{Magana2012, Matos2009, Suarez:2013iw, Urena-Lopez:2019kud}, 
they still present some open issues, for example, for scalar field dark 
matter (SFDM) numerical simulations have shown that the mass of the field could vary for different 
scales of the simulation in order to fit the observations \cite{Mocz:2019pyf}; while for 
dark energy, a single scalar field is not able to cross the PDL as shown in several results \cite{Cai:2009zp}.
Hence, the inclusion of more than a single field has come to the rescue.
Other areas have come up with similar ideas where two or more fields are used, for instance, a combination
of the inflaton and the SFDM \cite{Padilla:2019fju}, the inflaton and the curvaton \cite{Benisty:2018fja}, 
two scalar fields to account for inflation \cite{Bamba:2015uxa, vazquez2018inflationary}, to dark 
energy \cite{Vazquez:2023kyx} or to dark matter \cite{Tellez-Tovar:2021mge}, 
interactions between dark energy and dark matter \cite{bertolami2012two}.
\\

Until now, neither cosmology nor particle physics has provided a definitive theory to describe the DM or DE. 
In this work we open up the possibility that both dark matter and dark energy may be composed of 
scalar fields, with different potentials, and with the addition of a non-minimal interacting term, we called
it Scalar Field Interacting Early Dark Energy (SF-IEDE).
In this regard, there are different methods to introduce the coupling, one of which makes use of the 
variational approach, adding an interaction Lagrangian~\cite{Boehmer:2015kta, Boehmer:2015sha} that 
couples dark matter as a fluid and dark energy as quintessence.
Perhaps the most commonly used approach is to introduce the coupling term at the level of 
conservation equations \cite{Potter:2011nv, DiValentino:2019jae, Escamilla:2023shf}. In this case, 
the coupling term could exist when both dark components are modeled as fluids \cite{Yang:2022csz}, or 
a quintessence field coupled to dark matter~\cite{Amendola:1999er, Kase:2019veo, Perez:2021cvg}, and, 
in our case, when both are scalar fields.
However, it is important to emphasize that at some point of the formalism, most of the authors 
replace the scalar field dark matter with a perfect fluid \cite{Costa:2014pba, An:2018vzw}, and most 
of them only focus on the late dark energy contribution.
However, there are also proposals for interacting EDE scenarios~\cite{Goh:2023mau,Gomez-Valent:2022bku} 
as well.
The novelty of this work is that we maintain both components as scalar fields along with the interacting
term. For scalar field DM, we utilize the quadratic and axionlike potentials, the latter referred 
to as trigonometric along this work, while for scalar field EDE, we use the AS potential; the 
interacting term is explained further below. 
\\

The paper is organized as follows. In Sec.\ref{sec:description} we give a brief overview of the 
relevant equations for the interacting formalism, at the background and linear perturbation level, we focus on scalar fields. 
In Sec.\ref{sec:non_interacting}, we present the minimally coupled interacting case,  
whereas in Sec.~\ref{sec:interacting} we explore the consequences of a non-minimally coupling, 
and finally in Sec.\ref{sec:conclusions} the main conclusions of the paper are given.
\end{section}

\section{Interacting scalar fields at the background and linear levels} 
\label{sec:description}

Throughout this work, we consider the dark sector consisting of two canonical scalar fields: 
one representing early dark energy (quintessence) and the other representing dark matter (SFDM).
Each of these fields is coupled to the other components of the universe only through gravity, 
meaning that their conservation equations are not modified by the presence of the fields. 
However, for the dark sector, a coupling term is introduced at the level of its conservation 
equations \cite{Valiviita:2008iv}, which allows the exchange of energy and momentum between 
the scalar fields, and additionally guarantees that the total energy-momentum tensor remains conserved.

\subsection{Effective coupling}

The equations of motion for the background evolution, considering a flat, homogeneous and 
isotropic universe, are given by
\begin{subequations}
    \label{eq:density-equations}
    \begin{eqnarray}
        \dot{\rho}_\psi &=& - 3H (\rho_\psi + p_\psi) + Q_\psi \, , \\
        \dot{\rho}_\phi &=& - 3H (\rho_\phi + p_\phi) + Q_\phi \, .
    \end{eqnarray}
\end{subequations}
Here, overdot means derivative with respect to cosmic time, $H$ is the Hubble factor, $\rho_\psi$ 
($p_\psi$) and $\rho_\phi$ ($p_\phi$) are the energy densities (isotropic pressures) of the fields 
$\psi$ and $\phi$, respectively. Furthermore, $Q_\psi$ ($Q_\phi$) is the rate of energy transfer to 
the DE component $\psi$ (the DM component $\phi$), and due to energy conservation, we find that 
$Q_\phi = - Q_\psi$. The form of the decay rates of the fields in the dark sector should be derived 
from first principles, but in the absence of a fundamental theory we have to resort to phenomenological 
proposals, and some may be more justified than others, see for instance~\cite{Majerotto_2010} and 
references therein.
\\

At the fundamental level, both scalar fields evolve according to their Klein-Gordon (KG) equations 
at both background and linear orders. These equations follow from the conservation of the energy--momentum  
tensor of a scalar field, which takes the expression 
$T_{\mu \nu}^{(\phi_A)}=\partial_{\mu}\phi_A\partial_{\nu}\phi_A-g_{\mu\nu}\left[\frac{1}{2}\partial^{\alpha}\phi_A\partial_{\alpha}\phi_A +V(\phi_A)\right]$. 
In correspondence with the equations of motion~\eqref{eq:density-equations}, at the background level, we find
\begin{subequations}
    \label{eq:Klein-Gordon}
    \begin{eqnarray}
        \ddot{\psi} + 3H \dot{\psi} +\partial_{\psi} V(\psi) &=& - \frac{Q}{\dot{\psi}} \, , \label{eq:Klein-Gordon-a} \\
        \ddot{\phi} + 3H \dot{\phi} +\partial_{\phi} V(\phi) &=& \frac{Q}{\dot{\phi}} \, . \label{eq:Klein-Gordon-b}
    \end{eqnarray}
\end{subequations}
Here, $Q = -Q_\psi = Q_\phi$ is an arbitrary function.  Note that the sign of the interaction determines the transfer of energy from one component to the other, and for our choice the energy flux goes from the field $\psi$ to the field $\phi$. If there is no interaction, $Q=0$, and then each scalar field evolves independently. 
 
\subsection{Equations of motion}

The strategy for solving the coupled KG Eqs.~(\ref{eq:Klein-Gordon}) at zeroth order is as follows:
given the rapid oscillations regime in the evolution of SFDM, we convert its associated KG equation 
into a system of first-order differential equations by introducing the following change of 
variables~\cite{Urena-Lopez:2015gur, Cedeno:2017sou}: 
\begin{eqnarray}
  \Omega^{1/2}_{\phi}\sin\left(\frac{\theta}{2}\right)=\frac{\kappa \dot{\phi} }{\sqrt{6}H}, \quad \Omega^{1/2}_{\phi}\cos\left(\frac{\theta}{2}\right)=\frac{\kappa V^{1/2}}{\sqrt{3}H}, \nonumber \\ 
   y_{1}=-\frac{2}{H}\sqrt{2}\partial_{\phi}V^{1/2}, \quad y_{2}=-\frac{4\sqrt{3}}{H\kappa}\partial^2_{\phi}V^{1/2}\,,
\end{eqnarray}

\noindent
where, $\kappa^2\equiv 8\pi G$, with $G$ the gravitational coupling
constant, and $\Omega_{\phi}=\kappa^2\rho_{\phi}/3H^2$. This change in variables has been shown to
accurately track the evolution of the SFDM system.
Applying this change of variables, and analogous to the results in \cite{Roy:2023uhc}, 
the associated system to Eq.(\ref{eq:Klein-Gordon-b}) is:
\begin{eqnarray}
   \dot{\theta} &=& -3 H \sin \theta +Hy_1 -\frac{\kappa^2 \cos \left(\frac{\theta}{2}\right)}{3H^2 \Omega_{\phi} \sin \left(\frac{\theta}{2}\right) }Q_{\phi} \,, \nonumber \\ 
   \dot{y_1}    &=& \frac{3}{2}H(1+w_T)y_{1}+\Omega_{\phi}^{1/2}\sin \left(\frac{\theta}{2}\right)y_{2}H \, , \nonumber \\ 
   \dot{\Omega}_{\phi} &=& 3H\Omega_{\phi}(w_T-w_{\phi})-\frac{\kappa^2}{3H^2}Q_{\phi} \, ,
\end{eqnarray}
where $w_T$ corresponds to the total equation of state (EoS), and $w_{\phi}$ is the EoS associated to the SFDM.
However, for the quintessence component and considering its slow evolution, we will preserve 
the original variables $\psi$ and $\dot{\psi}$.
\\

At linear order, in Fourier space and in synchronous gauge~\cite{Ma:1995ey}, the evolution of each 
scalar field is dictated by the perturbed KG equation:
\begin{subequations}
\label{eq:linear_KG}
    \begin{eqnarray}
       \ddot{\delta \psi} +3H\dot{\delta \psi} +\left[k^2 +\partial^2_\psi V(\psi) \right] \delta \psi + \frac{1}{2} \dot{h} \dot{\psi}=- \delta \left(\frac{Q}{\dot{\psi}}\right) ,~ \label{eq:linear_KG-a} \\
       \ddot{\delta \phi} +3H\dot{\delta \phi} +\left[k^2 +\partial^2_\phi V(\phi) \right] \delta \phi + \frac{1}{2} \dot{h} \dot{\phi}=\delta \left(\frac{Q}{\dot{\phi}}\right) ,~~ \label{eq:linear_KG-b}
    \end{eqnarray}
\end{subequations}
where $\delta \psi$ and $\delta \phi$ are the perturbations of the scalar fields, and $h$ is the 
trace of the metric perturbation. 
    
To solve the equations, we follow the same line of thought as in the background case.
Quintessence will be implemented through Eq. (\ref{eq:linear_KG-a}) and for SFDM, i.e. Eq. \eqref{eq:linear_KG-b}, we will make use of angular variables,
\begin{eqnarray}
    &&\frac{2}{3}\frac{\kappa\dot{\delta \phi}}{H}=-\Omega_{\phi}^{1/2}e^{\alpha}\cos(\vartheta/2)\,,  \frac{\kappa y_{1}\delta \phi}{\sqrt{6}}=-\Omega_{\phi}^{1/2}e^{\alpha}\sin(\vartheta/2)\,,\nonumber \\
    &&\delta_{0}=-e^{\alpha}\sin[(\theta-\vartheta)/2]\,,\delta_{1}=-e^{\alpha}\cos[(\theta-\vartheta)/2]\,,
\end{eqnarray}
then the evolution of Eq. \eqref{eq:linear_KG-b} is given in terms of $\delta_0$ and $\delta_1$, as follows:    
\begin{eqnarray}
    \dot{\delta}_0 &=&H\left[-3\sin \theta -\omega(1-\cos \theta)\right]\delta_{1}+H\omega \sin \theta \delta_{0} \nonumber \\
    &-&\frac{\dot{h}}{2}(1-\cos \theta) 
    -\sqrt{\frac{2}{3}}\frac{\kappa}{\Omega_{\phi}^{1/2}}\sin\left(\frac{\theta}{2}\right)\delta\left(\frac{Q}{\dot{\phi}}\right) \nonumber \\
    &+& \frac{\kappa^2 Q}{3H\Omega_{\phi}}\left[\frac{\delta_{0}}{2}-\cot\left(\frac{\theta}{2}\right)\delta_{1} \right] \, , \\
    \dot{\delta}_1 &=&\left[-2\cos \theta -\omega \sin \theta +\Omega_{\phi}^{1/2}\sin\left(\frac{\theta}{2}\right)\frac{y_2}{y_1}\right]H\delta_{1}    \nonumber \\
    &-& \frac{\dot{h}}{2}\sin \theta+ \left[\omega(1+\cos \theta)-\Omega_{\phi}^{1/2}\cos\left(\frac{\theta}{2}\right)\frac{y_2}{y_1}\right]H\delta_{0}   \nonumber \\
    & -& \sqrt{\frac{2}{3}}\frac{\kappa}{\Omega_{\phi}^{1/2}}\cos\left(\frac{\theta}{2}\right)\delta\left(\frac{Q}{\dot{\phi}}\right) \nonumber \\
    &+& \frac{\kappa^2 Q}{3H\Omega_{\phi}}\left[\frac{\delta_{1}}{2}+\cot\left(\frac{\theta}{2}\right)\delta_{0} \right]\, .
\end{eqnarray}
 Here $\omega= k^2/k_{J}^2$, with the Jeans scale defined as $k_{J}^2 \equiv 2a^2H^2y_{1}$.   
\\

Up to this point, the formalism remains general, as we have not specified the quintessence or SFDM potential yet, nor have we assumed a specific expression for the interacting term. It is important to note that, to solve the background and linear equations, these specifics should be provided. 
In fact, for linear equations, it becomes necessary to substitute the function $Q$ and then perturb the terms $Q/\dot{\psi}$ and $Q/\dot{\phi}$ in Eqs.~\eqref{eq:linear_KG}. In the subsequent sections, we will provide examples of these functions to solve the system by modifying the Einstein--Boltzmann solver \textsc{class}\footnote{\url{https://github.com/gabygga/Interacting.git}}, to then obtain the cosmological observables.    

\section{Non interacting case}\label{sec:non_interacting}

In this section, as an initial step, we begin by summarizing the key individual characteristics of SFDM and EDE potentials, given that it is common to model the dark sector with only one component as a scalar field. Then, we will explore the scenario where both components are described simultaneously by scalar fields, considering as a first attempt that they are coupled solely through gravity, i.e., $Q=\delta Q=0$.

\begin{figure}
    \centering
    \includegraphics[width=0.48\textwidth]{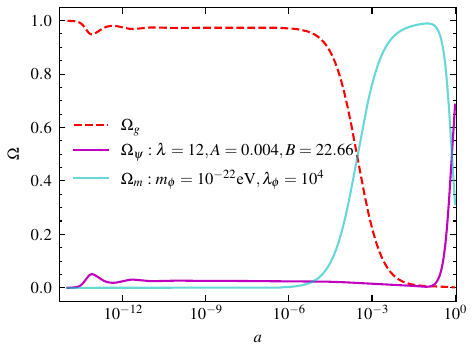}
    \caption{AS($\psi$)-SFDM($\phi$) model. Evolution of the density parameters, including radiation ($\Omega_{\rm g}$), matter ($\Omega_{\rm m}$: SFDM + baryons), and quintessence as dark energy ($\Omega_\psi$). The scalar field parameters are indicated by the labels. This parameter selection allows us to have a moderate, non-null contribution of the quintessence scalar field at early times, while the parameters of the SFDM are chosen for the trigonometric potential.}
    \label{fig:Omega_2sf}
\end{figure}
\begin{figure}
    \centering
    \includegraphics[width=0.48\textwidth]{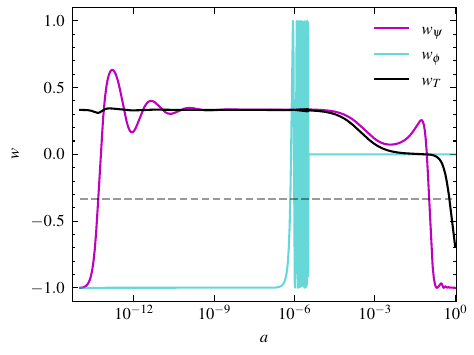}
    \caption{Evolution of the EoS of the scalar fields with the same parameters as in Fig. \ref{fig:Omega_2sf}. The gray dashed line corresponds to $-\frac{1}{3}$.  }
    \label{fig:w_2sf}    
\end{figure}

\subsection{Background}
At the background level, the evolution of each scalar field remains unaffected by the gravitational influence of the other. 
Due to this characteristic, in this subsection both scalar fields are turned on, {and the chosen scalar field parameters to sketch each component are such that at the present time become $\Omega_m=0.32$ and $\Omega_{\psi}=1-\Omega_{m}$.}
\subsubsection{Quintessence scalar field}

Pure exponential quintessence is known to not provide a transition from a matter-dominated epoch described by a scaling solution, where the contribution of the scalar field is not exactly zero, to an accelerated expansion epoch \cite{Tsujikawa:2013fta}. Therefore, if we want to have an early scalar field contribution, one approach is to modify the exponential potential. An option we opt to take is to consider the Albrecht-Skordis potential, which additionally has the property that cannot be classified as purely thawing or freezing, the mathematical expression used in this work is \cite{Albrecht:1999rm}:
\begin{equation}\label{eq:AS}
    V(\psi) = V_{\rm p}(\psi) e^{-\lambda \psi}=\left\{(\psi -B)^{2}+A\right\} e^{-\lambda \psi}\, , 
\end{equation}
where $B$, $A$ and $\lambda$ are constants. In order to achieve late-accelerated
expansion and a scaling solution at early times, these constants are not totally independent. 
For instance, $A\lambda^2<1$, and $B$ should be determined by other cosmological observables, see Appendix~\ref{appendix:ASpotential}
for more details of this potential.

Note that in recent work~\cite{Skordis:2000dz, Adil:2022hkj} the AS potential~\eqref{eq:AS} has
been rewritten. 
At the background level, our focus lies on the evolution of the density and EoS parameters, 
as depicted in Figs. \ref{fig:Omega_2sf} and~\ref{fig:w_2sf} respectively.
From Fig.~\ref{fig:Omega_2sf} we can see that, at early times, $\Omega_{\psi} \sim 0.028$ 
is non-negligible and that it is scaling with respect to radiation, as expected.
In Fig.~\ref{fig:w_2sf}, we show the EoS parameters of the scalar fields, and also the
effective one (defined as $w_{T}=P_{\rm tot}/\rho_{\rm tot}$), from this figure it is clear that 
for the AS potential, there is no need to include an additional cosmological constant, 
since the scalar field can eventually reach a value of $w_{\psi}=-1$ at late times and 
hence drives the accelerated expansion of the universe.

 \subsubsection{Scalar Field Dark Matter}
 
The formalism of this work is applicable to the following three different SFDM potentials that can be differentiated by a single 
parameter $\lambda_{\phi}$ \cite{Cedeno:2017sou}, related to the decay constant, $f$, 
for instance $\lambda_{\phi} \equiv \mp 3/\kappa^2 f^2$:
 \begin{equation}\label{eq:SFDM_potentials}
  V(\phi) = \left\{
    \begin{array}{@{\quad}l@{\quad}l@{}}
      m_{\phi}^2 f^2\left[1+\cos(\phi/f)\right]\,, & \text{ $\lambda_{\phi} >0$, } \\
      \frac{1}{2}m_{\phi}^2\phi ^2\,, & \text{ $\lambda_{\phi} =0$, }\\
      m_{\phi}^2 f^2\left[1+\cosh(\phi/f)\right]\,, & \text{ $\lambda_{\phi} <0$. }
    \end{array}
    \right.
\end{equation}
These potentials drive different background and perturbed density evolution.
In Figs.~\ref{fig:Omega_2sf} and~\ref{fig:w_2sf}, the evolution of the density and the 
EoS parameter are plotted for an SFDM mass of $10^{-22} \rm{eV}$ and for the trigonometric 
potential. These plots show that the field evolves similar to CDM but at early times, 
where its contribution to the total matter content is negligible, there is a period where it evolves similarly to a cosmological constant ($w_{\phi}\sim -1$) ~\cite{Hlozek:2014lca}.

\subsection{Linear Perturbations}
\subsubsection{Quintessence scalar field}

\begin{figure}
    \centering
    \includegraphics[width=0.48\textwidth]{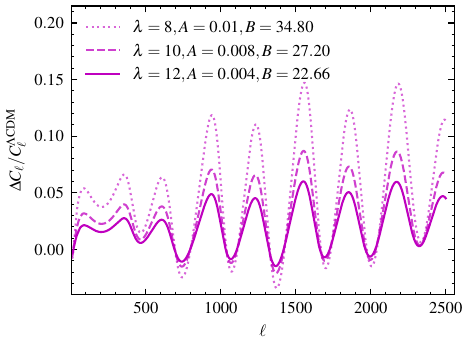}
    \includegraphics[width=0.48\textwidth]{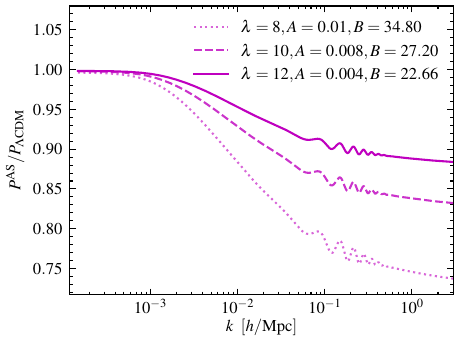}
    \caption{AS-CDM model. The upper panel is for residual CMB-TT and the bottom panel is for the ratio of MPS with respect to $\Lambda$CDM. The AS parameters of the different lines are indicated by the labels.}
    \label{fig:AS_MPS_TT}
\end{figure}

As is well known, the scalar field has linear perturbations such that if this field
contributes at early times, its perturbations are expected to result in modifications 
to the CMB and the matter power spectra (MPS). 
In order to isolate the effects of each field, in this part, we only consider EDE as a 
scalar field and DM as dust (AS+CDM). To illustrate these modifications, in Fig.~\ref{fig:AS_MPS_TT}  
we plot the residual CMB-TT spectra and the MPS ratio with respect to $\Lambda$CDM, 
along with three sets of AS parameters to demonstrate their impact.
The other baseline $\Lambda$CDM parameters are from Planck18 \cite{Planck:2018vyg} and 
are the same for all datasets, { with the following values: $\omega_b=0.022$, $\Omega_{\phi}=0.264$, $\tau=0.054$, $A_s=2.1\times 10^{-9}$, $n_s=0.966$, and $h=0.67$.}

Some general aspects are common to these spectra. A lower value of $\lambda$ corresponds
to a larger fraction of EDE during the radiation-dominated epoch, leading to significant 
suppression of matter perturbations during that period. 
This implies that smaller $\lambda$ values result in greater deviations from the 
reference model. In turn, this suppression of power in the MPS occurs on small scales 
(large $k$'s), while at smaller $k$ modes, the MPS associated with the AS model approaches 
$\Lambda$CDM, making the two practically indistinguishable on large scales. 
This suppression of matter perturbation growth translates into a weaker gravitational 
potential, leading to an enhanced CMB at large multipoles, as the potentials are not strong 
enough to capture the photons. Given the high precision of CMB spectra measurements, 
this could suggests that the AS-CDM model needs -at least- a larger matter contribution, {$\omega_m$} to provide 
a good fit to the data.
{By doing so, it avoids the reduction of the angular size of the sound horizon $\theta_s$, caused by EDE. An option to achieve this is to increase the $h$ value as the parameter $\lambda$ decreases. Another approach would be to keep $\theta_s$ fixed and then obtain the corresponding $h$ for each $\lambda$. In this case, to keep the sound horizon fixed, $\omega_m$ will increase as the EDE fraction does.}

\subsubsection{Scalar field dark matter}

\begin{figure}
    \centering
    \includegraphics[width=0.48\textwidth]{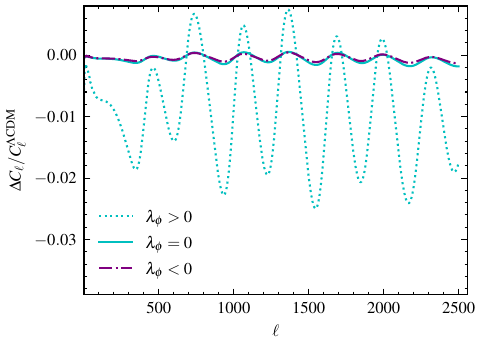}
    \includegraphics[width=0.48\textwidth]{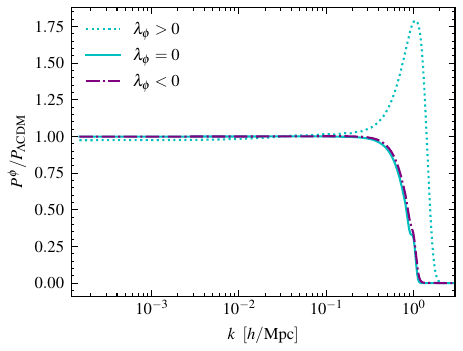}
    \caption{$\Lambda$-SFDM model. Cosmological spectra for the different SFDM potentials, indicated by the labels, all of them for a fixed mass,  $m_{\phi}=10^{-24} \rm eV$. The upper panel displays the residual CMB-TT spectra, while the bottom panel shows the MPS ratio relative to $\Lambda$CDM.}
    \label{fig:SFDM_spectra}
\end{figure}

In the main results, we will focus only on the quadratic and trigonometric potentials 
because the hyperbolic gives rise to very similar effects to those of the quadratic 
one~\cite{LinaresCedeno:2021sws}. However, the trigonometric potential exhibits a different 
behavior, especially in its corresponding MPS, which is significantly different around the 
cutoff scale, as illustrated in Fig.~\ref{fig:SFDM_spectra}. At the smallest scales, 
it produces a bump, which is relevant to this work because this behavior is also opposite 
to the EDE as shown in Fig.~\ref{fig:AS_MPS_TT}. 

To distinguish the linear effects of the SFDM from that of quintessence, in the 
plots of Fig.~\ref{fig:SFDM_spectra}, we use a cosmological constant as dark energy ($\Lambda$+SFDM), 
and to emphasize the effects for CMB and MPS we chose a smaller mass value rather 
than the selected for the background plots, {and the remaining relevant cosmological parameters were fixed to be the same than in the EDE case; $\omega_b=0.022$, $\Omega_{\phi}=0.264$, $\tau=0.054$, $A_s=2.1\times 10^{-9}$, $n_s=0.966$, and $h=0.67$.} Note that the CMB-TT remains largely unchanged 
when considering the hyperbolic and quadratic potentials, but there is a more pronounced 
deviation with the trigonometric potential. Regarding the MPS ratio, for the same parameter 
values, the most significant difference with respect to $\Lambda$ CDM occurs in large wave
numbers. However, it is important to note that, overall, the trigonometric 
potential results in a slightly smaller MPS at large scales.

\subsubsection{SFDM and Quintessence}

\begin{figure}
    \centering
    \includegraphics[width=0.48\textwidth]{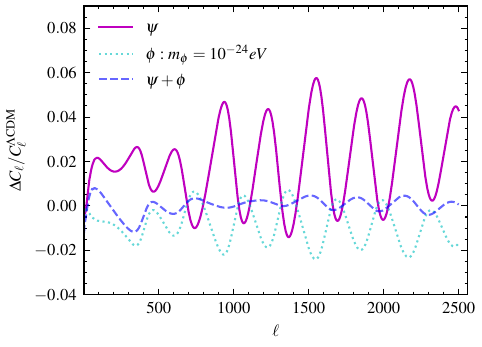}
    \includegraphics[width=0.48\textwidth]{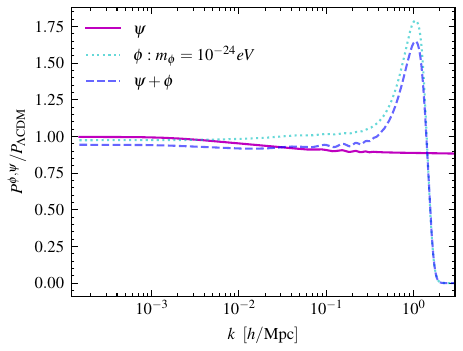}
    \caption{AS($\psi$)-SFDM($\phi$) model. The upper panel shows the residual CMB-TT plot, where $\Delta C_{\ell}=C_{\ell}^{\phi,\psi}-C_{\ell}^{\Lambda\rm{CDM}}$, while the bottom panel displays the MPS ratio with respect to $\Lambda$CDM. In these plots, we are using an SFDM mass of $10^{-24}\rm{eV}$, which differs from the mass used in the background plots, while the quintessence parameters remain the same as before, {and all of them have the same $\Omega_m=0.316$}. The dotted line assumes only DM as scalar field ($\phi$), the solid line corresponds to DE as quintessence ($\psi$), and the dashed line represents the scenario where dark matter and dark energy are described as scalar fields without interaction.}
    \label{fig:spectra_AS_trig}
\end{figure}

Considering both scalar fields activated,
it is evident from the upper panel of Fig.~\ref{fig:spectra_AS_trig}, that the deviations caused by the 
EDE are partially compensated for by those induced by the scalar field dark 
matter. For example, the plots illustrate that in this scenario, the deviations caused by EDE 
in the CMB-TT spectrum are {partially} counteracted by those generated by SFDM, resulting in a more 
moderate overall deviation. {For instance the angular sound horizon for the considered cosmological parameters is $100\theta_s=1.041$, for the cosmological standard model, whereas in the  AS$(\psi)$-SFDM$(\phi)$ model it changes to  $100\theta_s=1.044$ for $\Lambda+$SFDM, $100\theta_s=1.033$ for AS+CDM and $100\theta_s=1.039$ when both scalar fields are present.\footnote{{The reported angular scale by the Planck collaboration~\cite{Planck:2018vyg} is $100\theta_s=1.0411 \pm 0.0003$. Given the resultant values above for each model, we see that CMB observations in general are sensitive to the different scenarios, under the same numbers of the cosmological parameters, and could then be used, in a full Bayesian analysis, to put tighter constraints on these possibilities.}}}
However, the bottom panel of Fig.~\ref{fig:spectra_AS_trig} displays the ratio between the MPS 
of the scalar fields and the corresponding to $\Lambda$CDM. When considering the 
trigonometric potential for the description of dark matter, the combined effect prevents the 
suppression of power caused by the EDE, but its characteristic pattern of acoustic 
oscillations is still present, while simultaneously reducing the size of the bump generated by the non--linearities of the trigonometric potential.

It is worth recalling that even in the nondirect dark-coupling case, the gravitational 
influence of both scalar fields is capable to modify the cosmological observables. This is 
in agreement with the evolution at the background and linear levels of the scalar fields. 
In particular, the linear contrast density will be discussed in the following section.

\section{Interacting case}\label{sec:interacting}

In this section, we turn to the case where a non-minimal coupling between the scalar fields is 
introduced, and it is added at the level of the KG equations~(\ref{eq:Klein-Gordon}), (\ref{eq:linear_KG}). 
Our equations can incorporate interacting terms that may arise from alternative formalisms, as elaborated in the Appendix~\ref{App:alternative}.
However, as an illustrative example 
we focus on a 
coupling that is proportional to the temporal derivatives of the scalar fields~\footnote{This 
coupling has the advantage of washing out the possible divergences carried by the ratios 
in Eqs. \ref{eq:Klein-Gordon} and its corresponding terms when angular variables are introduced. 
Notice that within the code the interaction kernel could be easily modifiable.}:
\begin{equation}\label{eq:interacting}
    Q=\beta \dot{\phi}\dot{\psi}\, ,
\end{equation}
where $\beta$ is the coupling constant (in units of $H_0$),  the convention adopted in 
this work is  that $Q, ~\delta Q>0$ means that SFDM is transferring energy density to the 
EDE component and,  it is important to remark that in this context also the inverse 
process is possible.
A more comprehensive and formal derivation regarding the introduction of the interaction term can be found in the Appendix~\ref{App:fluid-approach}.
\\

\begin{figure}
    \centering
    \includegraphics[width=.48\textwidth]{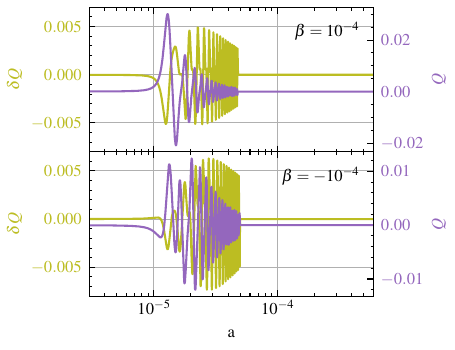}
    \caption{Interacting $Q-\psi\phi$ model. Evolution of the interacting term according to Eq.~\ref{eq:interacting}. The upper panel corresponds to $\beta > 0$, while the bottom panel illustrates the scenario with the opposite sign of $\beta$. In both panels, the purple color is for the background level (right vertical axis) and the olive color for linear perturbations (left vertical axis).}
    \label{fig:interaction}
\end{figure}

\begin{figure}
    \centering
    \includegraphics[width=0.48\textwidth]{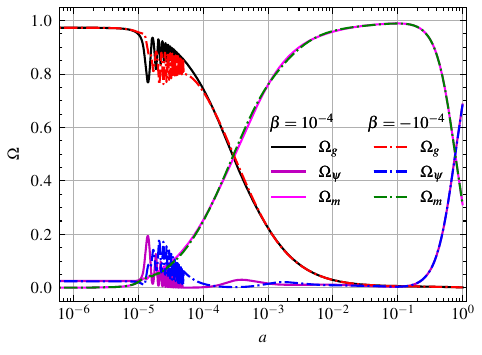}
    \caption{Evolution of the matter densities when the direct interaction in the dark sector is turned on, with $\beta=\pm10^{-4}$. Quintessence is described by the AS potential with parameters: $\lambda=12, A=0.004, B=22.66$, while $\Omega_{m}=\Omega_{b}+\Omega_{\phi}$ with SFDM parameters: $m_{\phi}=10^{-24}\rm{eV}, \lambda_{\phi}=10^4  $. Solid lines correspond to the $\beta >0$ case and dashed--dotted lines are for the $\beta<0$ case. Regardless of the sign of $\beta$, the interaction causes EDE to capture the oscillations originally present in the SFDM component.}
    \label{fig:beta}
\end{figure}

Just like in the case without a direct interaction, in this section, the EDE is 
represented by a quintessence scalar field with an AS potential given by Eq.~\eqref{eq:AS}, 
and the SFDM can be described by the potentials presented in Eq.~(\ref{eq:SFDM_potentials}) 
(model $Q-\psi\phi$) {and the cosmological parameters are equal to those in previous sections}.
At the background and linear level, the evolution of the matter densities is sensitive
to the value of the constant coupling.
For example, for a fixed SFDM mass of $10^{-24}\rm{eV}$, $|\beta|$ values lower than $10^{-4}$ 
produce smaller amplitudes or changes less pronounced than those presented in 
Figs.~(\ref{fig:interaction}, \ref{fig:beta}). 
However, the overall shape remains similar, showing a rescaled pattern.
Also, if we consider SFDM masses higher than $10^{-24}\rm{eV}$, the interacting effect 
is barely noticeable for the same value of $\beta$. 

For instance, the interaction rate is non-negligible over a range of the scale factor, 
shown in Fig.~\ref{fig:interaction}. This is a consequence in agreement with the proposal 
that the interacting term is proportional to the kinetic terms of the scalar fields, 
see Appendix \ref{App:unidirectional} for more details. 
Specifically, this has to do with the fact that the SFDM starts oscillating rapidly, 
with a mean - on average - equal to zero.
It is also noticeable that the oscillations of the interaction kernel are changing sign, 
in both cases of the sign of $\beta$, suggesting a bidirectional exchange between SFDM and EDE. 
A similar change in sign in the interacting kernel, at low redshift, has been found using model-independent techniques \cite{Escamilla:2023shf}.
The difference between the upper and lower panels of Fig.~\ref{fig:interaction} is that 
for $\beta<0$ the background interaction has a small reduction in amplitude, while at linear order the amplitudes of $\delta Q$ are enhanced.

In Fig.~\ref{fig:beta}, we can see the effects of the density interchange between 
SFDM and EDE. It is noticeable that now, where the interaction is relevant, the EDE has an 
extra contribution ceded by the kinetic term of the SFDM, that is reduced for negative 
values of $\beta$. It is also worth noting that the effect we are observing in the 
radiation component does not indicate that radiation is providing density to EDE. 
Instead, this is a consequence of the scaling behavior of EDE with respect to radiation.
In both cases of Fig.~\ref{fig:beta}, the interaction is diluting while oscillating, and, 
notably, the global shapes of $\Omega_{\psi}$ exhibit similar patterns. For both signs of 
$\beta$, during the oscillating phase of the interaction, there is a reflected oscillating 
enhancement of this component (slightly smaller for $\beta<0$), which then dilutes as 
the interaction progresses.
Subsequently, when the interaction stops, the fields evolve freely and its standard 
evolution, $\beta=0$, is recovered.
\\

\begin{figure}
    \centering
    \includegraphics[width=0.48\textwidth]{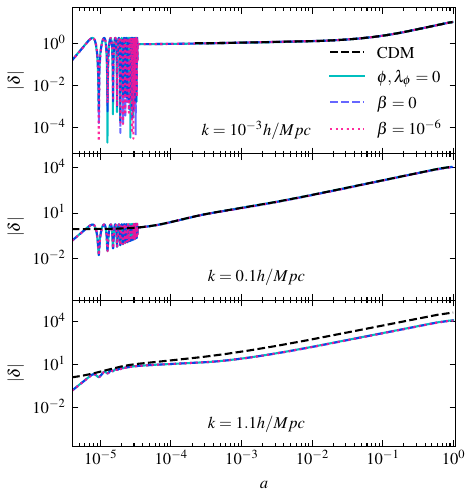}
    \caption{Evolution of different contrast densities for a quadratic SFDM potential. For the different models the mass of the SFDM is fixed to $10^{-24}\rm eV$ and the AS parameters are $\lambda=12, A=0.004, B=22.7$. }
    \label{fig:contrasts_quad}
\end{figure}

\begin{figure}
    \centering
    \includegraphics[width=0.48\textwidth]{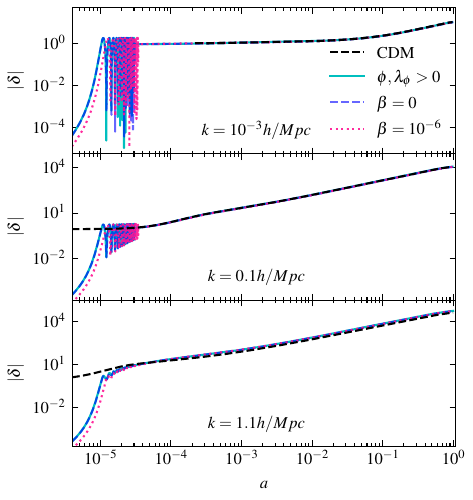}
    \caption{Evolution of different fractional density contrasts. For the different models the mass of the SFDM is fixed to $10^{-24}\rm eV$ with a trigonometric potential, $\lambda_{\phi}=10^4$. The AS parameters are $\lambda=12, A=0.004, B=22.7$. The evolution with a negative $\beta$ value has slightly pronounced and prolonged effects than the $\beta>0$ scenario.}
    \label{fig:contrasts}
\end{figure}

At the level of linear perturbations, we are able to gain some insights into the effect 
of the interaction by looking at the evolution  of the dark matter fractional density contrasts, 
$\delta=\delta \rho_{\rm dm}/\rho_{\rm dm}$. 
In Figs.~\ref{fig:contrasts_quad} and~\ref{fig:contrasts}, we present the evolution of 
three distinct $k$ modes under different assumptions. For comparison, we include the 
corresponding evolution of CDM and a case with only dark matter as a scalar field. We will 
give more emphasis to the trigonometric potential, but a detailed analysis on the quadratic 
potential can be found in \cite{Hlozek:2014lca}.

The mode $k=10^{-3}h/\rm{Mpc}$ oscillates very rapidly until the scale factor reaches 
a value $a\sim 10^{-5}$, long before it enters the horizon. However, when it crosses the 
horizon, it grows as CDM, although its strength is slightly lower. At this scale, there 
is no significant difference when both scalar fields are present, even for the 
interaction parameter $\beta=10^{-6}$. However, this could change for different values 
of the interaction parameter $\beta$. 
%
From the evolution of the intermediate wave number $k=0.1 h/$Mpc, we notice that it 
grows very similarly in the SFDM case as in the $\Lambda$CDM case, except for earlier times 
when the density contrast of the SFDM initially experiences suppressed growth.  However, 
it follows the same scaling with a period of rapid oscillations, and eventually scales like 
the CDM case at late times. The initial suppression is more pronounced for the 
trigonometric potential than for the quadratic potential. During this period, the 
interaction term exhibits some differences in growth.
Finally, the scale at the wavenumber $k=1.1 h/$Mpc enters the horizon before the 
matter-radiation equality, and it is beyond the characteristic Jeans scale.  From the evolution 
of this and the wave numbers beyond the characteristic cut-off, as shown in Fig.~\ref{fig:contrasts_quad}, 
we can see that for the quadratic potential, the evolution of dark matter (DM) is lower than 
that of cold dark matter (CDM). Consequently, we observe a suppression of power at these scales.  
However, for the trigonometric potential, the evolution of the density contrast DM 
in Fig.~\ref{fig:contrasts} is not below that of the corresponding CDM; instead, it is 
slightly larger, resulting in the characteristic bump in the MPS for this potential.
\\


\begin{figure}
    \centering
    \includegraphics[width=0.48\textwidth]{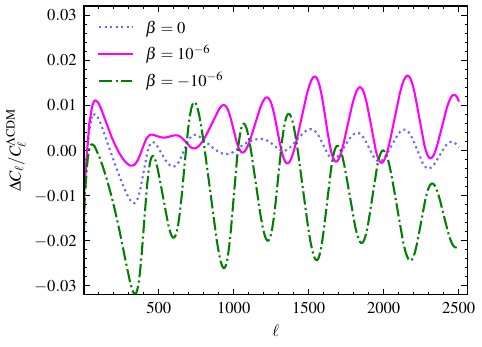}
    \includegraphics[width=0.48\textwidth]{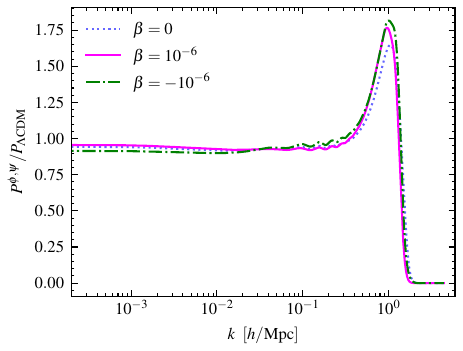}
    \caption{Upper panel shows the CMB-TT spectra and the bottom exhibits the MPS, both described by the same cosmological and scalar field parameters.
    The interaction between the scalar fields is considered for different values of $\beta$ as indicated by the labels. The values of the scalar fields parameters are the same as in Fig.~\ref{fig:contrasts}
    }
    \label{fig:spectra_int}
\end{figure}

Fig.~\ref{fig:spectra_int} shows the changes in the CMB-TT and matter power spectra for 
different values of the interaction parameter and considering the trigonometric potential 
to describe the SFDM component. 
Concerning the matter power spectra, we observe that when the interaction is considered 
and activated with different values of $\beta$, the overall shape of the total spectra closely 
mimics that of the SFDM. Nevertheless, it also reflects the oscillations of the MPS associated 
with EDE at intermediate--small scales.

The part of the spectra that corresponds to the early universe (large $k$ modes) reflects 
that for $\beta\neq0$ the non--linearity's associated to the trigonometric potential are 
not erased by the interaction, instead, they are enhanced regardless of the sign of 
the interacting parameter. However, the difference is that $\beta>0$ transfers more of 
its kinetic term to EDE than in the case $\beta<0$, so that the spectra of positive $\beta$ 
are slightly smaller than the opposite. 
At intermediate-small scales, where the EDE starts to suppress the spectra, this suppression 
does not persist to larger $k$ modes as it does in the case of EDE alone. 
Instead, this suppression is counteracted by the enhancement caused by SFDM.

Finally, on intermediate--large and large scales ($k\lesssim 0.03[h/{\rm Mpc}]$), 
Fig.~\ref{fig:spectra_int}  shows  that  $\beta<0$ results in a suppression of power. 
This might seem counterintuitive, but it is a consequence that these scales bring the 
horizon around the time when the kinetic term of the EDE field is recovering 
$(a\gtrsim 1.7\times 10^{-3})$ and, indeed, its density is slightly greater than that 
corresponding to the $\beta=0$ case. 
By the same epochs, the $\beta>0$ case experiences an opposite behavior, resulting now 
in a decrease in EDE at these wavenumbers. This makes these transients reflect at the 
point where the spectra cross.

In relation to the impact on the CMB-TT power spectra, as illustrated in the upper panel 
of Fig.~\ref{fig:spectra_int}, it is important to remark that the net effect in the residual 
plots for the chosen parameters is about 3\%, compared with the $\Lambda$CDM model, {however it is important to remark that these deviations could be different if the cosmological parameters were fixed using other basis}. 
Notably, these deviations are more pronounced in the case of $\beta<0$, where the same $|\beta|$ 
leads to enhanced perturbations in the considered model of interaction, however 
the smaller deviations are for the case of nondirect coupling $(\beta=0)$.
In alignment with the shape of the MPS, it is observed that the TT spectra now exhibit 
opposite peaks, suggesting that the corresponding densities lead, {on one side,} to potential wells with 
contrary behavior, as expected, {while on the other hand they also cause opposite impact on the angular sound horizon. Consequently, the deviations in the CMB-TT spectra naturally differ.}

\section{Conclusions}\label{sec:conclusions}
In this work we considered the possibility that both, simultaneously, the dark matter and 
dark energy can be described by scalar fields each with an associated potential that is 
able to reproduce the desired behavior, and, on the top of that, incorporating an 
interaction kernel between them that is consistent with the total conservation equations. 
This consideration leaves a footprint in the cosmological observables as shown 
in Figs.~\ref{fig:spectra_AS_trig} and~\ref{fig:spectra_int}, where the MPS and CMB 
power spectra are plotted for different cases: only one scalar field, two scalar 
fields coupled only by gravity, and an example of a non-minimal coupling 
(Eq.~\ref{eq:interacting}) controlled by an interacting kernel $Q$.

At the background level, when considering both scalar fields, the evolution of their 
associated matter densities is not altered by the gravitational presence of the other 
scalar field. However, when a direct interaction is activated, they exchange density; 
specifically for the chosen interaction, this exchange occurs through an interchange of 
kinetic energy, whose direction along the scale factor does not always correspond to 
the sign of $\beta$.
For example, for a fixed value of $\beta$, SFDM does not always yield/gain a part of 
its density. Instead, during certain periods of time, it receives/losses a contribution
coming from EDE. When considering two values of $\beta$ with the same absolute value but 
with a change in sign, we observe similar behaviors with a slight decreasing amplitude of 
the effects carried by $\beta<0$

The fluctuations in the background and linear densities caused by each one of the 
scalar fields imprint characteristic patterns in the cosmological spectra, which can 
be partially compensated by the direct or non-direct interaction between them. For example, 
the choice of the SFDM potential results in an MPS with a dramatic suppression of power at 
large $k$--modes if the considered potential is quadratic or hyperbolic. In contrast, 
it exhibits a bump if the potential is trigonometric.
 
In this respect, we found that the presence of an EDE scalar field component, described by 
the AS potential, can partially counteract these effects, as shown in
Figs.~\ref{fig:spectra_AS_trig} and~\ref{fig:spectra_int}. However, at these scales, 
the effects of SFDM are predominant over those from EDE, and thus the mechanism is not 
enough to entirely prevent the dramatic suppression of the MPS or eliminate the bump, whether 
there is a direct or non-direct interaction between the fields. Moreover, we find that the 
most significant attenuation occurs when the coupling is only gravitational, in contrast 
to the interaction described by the term~\ref{eq:interacting}, which, for example, 
increases the nonlinearities.

At large and intermediate scales the main characteristics of the SFDM are preserved, but now, 
even the global shape is unchanged, we expect small deviations with respect to $\Lambda$CDM 
depending the value of $\beta$ carried out by the over/underdensities created by the exchange 
of the kinetic term between the fields, as already explained in the text, this behavior is 
opposite for the same magnitude of the interaction but with different sign. 

Finally, concerning the MPS, it is also important to mention that now the suppression of 
power associated to the presence of a pure EDE field is avoided, 
this is relevant because it is common to compensate for this suppression by adding 
extra {$\omega_c$}.
The addition of this {component} is required to not alter other observables, for example, without 
it the potential wells {and the angular sound horizon decreases}, causing an enhanced CMB-TT 
spectrum. In that sense, a very important result is that in this scenario, {may be no} needs to add additional $\omega_{c}$ to have CMB-TT power spectra whose deviations could be in agreement 
with the data, for example, for the chosen parameters, we obtain deviations up to 3\% that 
can be lowered by different values of $\beta$. 
\\

Performing a Bayesian analysis to constrain the parameters of the interacting model 
along with those from the cosmological standard model is one of the future perspectives 
of this research. Up to this point, this study offers valuable insights into the intricate 
dynamics of the cosmological dark sector described by different scalar fields, and also 
provides a general formalism to implement different couplings between them, which can help 
to overcome some of the disadvantages associated with the models of a single scalar field.
Furthermore, the specific interaction discussed has been incorporated into a publicly 
available modification of a \textsc{class} code, enabling the easy inclusion of 
different couplings to expand the scope of the analysis concerning interacting scalar fields.

\section*{Acknowledgments}
GG-A acknowledges CONAHCYT postdoctoral fellowship and the support of the ICF-UNAM. LAU-L acknowledges partial support from the Programa para el Desarrollo Profesional Docente; Direcci\'on de Apoyo a la Investigaci\'on y al Posgrado, Universidad de Guanajuato; CONACyT M\'exico under Grants No. 286897, No. 297771, No. 304001; and the Instituto Avanzado de Cosmolog\'ia Collaboration. JAV acknowledges the support provided by FOSEC SEP-CONACYT Investigaci\'on B\'asica A1-S-21925, UNAM-DGAPA-PAPIIT IN117723 and FORDECYT-PRONACES-CONACYT/304001/2019.

\appendix
\section{Analysis of the AS quintessence potential}\label{appendix:ASpotential}
The standard parameterization of the AS potential is a bit confusing, so for a better
understanding of its intrinsic properties, we will consider the following changes. First, we make a change in the scalar field: $\psi -B \to \psi$, which leaves the KG equation of
motion unchanged. Secondly, we write the AS potential as
\begin{equation}\label{eq:AS_new}
    V(\psi) = \left( \mu^2 f^2 + \frac{1}{2} \mu^2 \psi^2 \right) e^{-\lambda \kappa \psi} \, , 
\end{equation}
where the new parameters $\mu$ and $f$ have the dimensions of energy, and $\lambda$ 
is now explicitly dimensionless in the exponential term. Note that the polynomial part
of the potential~\eqref{eq:AS_new} is parameterized as usual for these kinds of potentials.

The calculation of the critical points $\psi_c$ of the potential~\eqref{eq:AS_new} is quite 
straightforward, and from condition $V^\prime (\psi_c) =0$ we obtain
\begin{equation}
    \lambda \kappa \psi_c = 1 \pm \sqrt{1-2\lambda^2 \kappa^2 f^2} \, . \label{eq:critical-a}
\end{equation}
Then, the second derivative at the critical points is
\begin{equation}
     V^{\prime \prime} (\psi_c) = \mp \sqrt{1-2\lambda^2 \kappa^2 f^2} \, \mu^2 e^{-\lambda \kappa \psi_c} \, ,  \label{eq:critical-b}
\end{equation}
which means that the critical point $\psi_c$ corresponding to the minus (plus) sign 
in Eq.~\eqref{eq:critical-a} is a minimum (maximum). 

Another important result is the value of the potential~\eqref{eq:AS_new} at the minimum, 
as this value would be the effective cosmological constant that the model provides once
the field evolves towards its critical value. The result is
\begin{equation}
    \Lambda_{eff} = V(\psi_c) = \frac{\mu^2}{\lambda \kappa} \psi_c e^{-\lambda \kappa \psi_c} \, . \label{eq:cosmo-eff}
\end{equation}
Being a minimum, the scalar field oscillates rapidly as it settles down into this critical point.

We can see the advantages of the new parametrization in Eq.~\eqref{eq:AS_new}: the 
parameter $\mu$ plays the role of a bare mass of the scalar field $\psi$~\footnote{Recall 
that the effective mass of the field is given by $m^2_{\rm eff} = V^{\prime \prime} (\psi_c)$, 
and then $\mu^2_{\rm eff} \propto \mu^2$.}, while $f$ is a new energy scale that determines 
the position of the critical points of the potential. With respect to the parameter $f$, we
can consider two extreme cases. The first is $f=0$, which results in the potential
\begin{equation}
    V(\psi) = \frac{1}{2} \mu^2 \psi^2 e^{-\lambda \kappa \psi} \, ,
\end{equation}
with critical points at the values $\lambda \kappa \psi_c = 1 \pm 1$. However, in this
case, there is no effective cosmological constant at late times, as we can see from
Eq.~\eqref{eq:cosmo-eff} that $\Lambda_{\rm eff} =0$.

Another case of interest is $2 \lambda \kappa f=1$, for which there is only one
critical value of the scalar field: $\lambda \kappa \psi_c = 1$, which is also an
inflection point. The field potential now reads
\begin{equation}
    V(\psi) = \left( \mu^2 f^2 + \frac{1}{2} \mu^2 \psi^2 \right) e^{-\lambda \kappa \psi} \, ,
\end{equation}
which means that the field continues rolling down the potential, but this time 
without oscillations at all.

\section{Alternative formalism for a coupling}\label{App:alternative}
\begin{subequations}
\begin{equation}
    \ddot{\psi}+ 3H \dot{\psi} +\partial_\psi V(\psi)= - \Gamma_\psi \dot{\psi} \, . \label{eq:KG-psi-a}
\end{equation}
Such a friction term is standard in models of inflation to study the decay of the 
inflation field into other particles, where $\Gamma$ represents the decay rate of the 
scalar field calculated from a given coupling. If we multiply Eq.~(\ref{eq:KG-psi-a}) by 
$\dot{\psi}$ on both sides, we can rewrite it in terms of energy density and pressure, 
as\footnote{A quick comparison with the standard formalism of coupled perfect fluids 
in cosmology suggests that the term on the right hand side of Eq.~\eqref{eq:KG-psi-b} has
the expected form in isentropic creation/ annihilation of particles, that is, the entropy 
per particle remains constant. Furthermore, we could identify $\Gamma_\psi$ as the rate of 
change of the number of particles in a comoving volume $a^3$. See~\cite{Zimdahl_1996,Zimdahl_2001} 
for a detailed discussion of the thermodynamics of two interacting cosmological fluids.}
\begin{equation}
   \dot{\rho}_\psi + 3 H (\rho_\psi + p_\psi) = Q_\psi = - \Gamma_\psi \dot{\psi}^2 \, . \label{eq:KG-psi-b}
\end{equation}
\end{subequations}

If we repeat the exercise for the DM field, we arrive to the counterpart of Eq.~\eqref{eq:KG-psi-a} 
for the field $\phi$,
\begin{subequations}
 \begin{equation}
    \ddot{\phi}+ 3H \dot{\phi} +\partial_\phi V(\phi)= \Gamma_\phi \dot{\phi} \, . \label{eq:KG-phi-a}
\end{equation}   
which in turn leads us to the counterpart of Eq.~\eqref{eq:KG-psi-b},
 \begin{equation}
    \dot{\rho}_\phi + 3 H (\rho_\phi + p_\phi) = Q_\phi = \Gamma_\phi \dot{\phi}^2 \, . \label{eq:KG-phi-b}
\end{equation}   
\end{subequations}

As mentioned above, the conservation of the total density, for the coupled DM-DE fields
in terms of $Q_\psi = - Q_\phi$, then requires $\Gamma_\phi \dot{\phi}^2 = 
\Gamma_\psi \dot{\psi}^2$. Motivated by arguments of symmetry between the interacting 
fields, we propose here that the coupling between the fields is of the form
\begin{subequations}
\begin{equation}
    Q_\phi = - Q_\psi = \Gamma \dot{\phi}^2 \dot{\psi}^2 \, , \label{eq:coupling-a}
\end{equation}
where $\Gamma$ would be a constant with the appropriate units. Another form of 
Eq.~\eqref{eq:coupling-a} can be found if we write the kinetic terms using the definitions 
of the density and pressure for each field, and then 
\begin{equation}
    \Gamma \dot{\phi}^2 \dot{\psi}^2 =  \Gamma (\rho_\phi + p_\phi) (\rho_\psi + p_\psi) \, . \label{eq:coupling-b}
\end{equation}
\end{subequations}

Equation~\eqref{eq:coupling-b} has an additional advantage, in addition to its
symmetrical form, which is that interactions between two fields should involve both energy
densities, so a quadratic form could be considered a more natural choice~\cite{B_hmer_2010}. 
For example, the decay rate of the field $\psi$ would be effectively given by $\Gamma_\psi = 
\Gamma \dot{\phi}^2$, and likewise for the decay rate of the field $\phi$: $\Gamma_\phi = 
\Gamma \dot{\psi}^2$. Furthermore, the sum of density and pressure in Eq.~\eqref{eq:coupling-b} 
also hints at the possibility that the coupling functions $Q_\phi, Q_\psi$ could be seen as
extra friction terms in the equations of motion~\eqref{eq:KG-psi-b} and~\eqref{eq:KG-phi-b}.

\section{The fluid approximation for the coupled dark sector}\label{App:fluid-approach}
 \begin{eqnarray}
    \nabla_{\nu}T^{\mu \nu}_\phi = \phi^{,\mu} \left[ \Box \phi - \partial_\phi V(\phi) \right] = Q^{\mu}_\phi \, , \\
    \nabla_{\nu}T^{\mu \nu}_\psi = \psi^{,\mu} \left[ \Box \psi - \partial_\psi V(\psi) \right] = Q^{\mu}_\psi \, ,
\end{eqnarray}
where $\Box$ is the Laplace-Beltrami operator, and the terms $Q^\mu$ are called the (covariant) 
energy-momentum transfer rates. We must impose the condition $Q^{\mu}_\phi = - Q^\mu_\psi$, 
so that the conservation of the joint energy is achieved by the two components in the form 
$\nabla_{\nu} (T^{\mu \nu}_\phi + T^{\mu \nu}_\psi) =0$. 

Generally, the coupling term for each field can be written as $Q^{\mu} = Q u^\mu + F^\mu$, 
where $u^\mu$ is the total four-velocity~\cite{V_liviita_2008}. We refer to $Q$ as the
parallel component, since $Q^\mu u_\mu = -Q$, and represents the transfer of energy, while
$F^\mu$ is the perpendicular component that satisfies the condition $F^\mu u_\mu =0$, 
and represents the transfer of momentum. For simplicity, in this study we shall
assume that only energy is exchanged between the dark components, i.e. $F^\mu=0$.

According to~\cite{Piattella_2014,Faraoni_2023}, the four-velocity of the scalar fields 
in the dark sector can be written as
\begin{equation}
    u^\mu_\psi = \frac{\psi^{,\mu}}{\sqrt{-\psi_{,\mu} \psi^{,\mu}}} \, , \quad u^\mu_\phi = \frac{\phi^{,\mu}}{\sqrt{-\phi_{,\mu} \phi^{,\mu}}} \, ,
\end{equation}
so that both of them comply with the normalization condition $u_\mu u^\mu = -1$. With these 
definitions, the KG equations of motion for the fields read
\begin{subequations}
\begin{eqnarray}
    \Box \phi - \partial_\phi V(\phi) &=& - \frac{\phi_{,\mu} Q^{\mu}_\phi}{\phi_{,\mu} \phi^{,\mu}} \, , \\
    \Box \psi - \partial_\psi V(\psi) &=& - \frac{\psi_{,\mu} Q^{\mu}_\psi}{\psi_{,\mu} \psi^{,\mu}} \, .
\end{eqnarray}    
\end{subequations}

It is then possible to recover the equations of motion~\eqref{eq:KG-psi-b} 
and~\eqref{eq:KG-phi-b} if we choose the following coupling terms for each field.
\begin{equation}
    Q^{\mu}_\phi = \Gamma_\phi \left( \phi_{,\mu} \phi^{,\mu} \right) u^\mu_\phi \, , \quad Q^{\mu}_\psi = \Gamma_\psi \left( \psi_{,\mu} \psi^{,\mu} \right) u^\mu_\psi \, . \label{eq:cov-coupling-a}
\end{equation}
Notice that Eqs.~\eqref{eq:cov-coupling-a} also gives more support to our choice of 
the phenomenological coupling in Eqs.~\eqref{eq:KG-psi-b} and~\eqref{eq:KG-phi-b}: 
the coupling is proportional in each case to an invariant quantity, which is the norm 
of the covariant derivative of the fields. More to the point, if we see the fields as a
joint dark sector, Eqs.~\eqref{eq:coupling-a} can be further written as
\begin{subequations}
\label{eq:cov-coupling-b}
\begin{eqnarray}
    Q^{\mu}_\phi &=& \Gamma \left( \phi_{,\mu} \phi^{,\mu} \right) \left( \psi_{,\mu} \psi^{,\mu} \right) u^\mu_\phi \, , \label{eq:cov-coupling-ba} \\
    Q^{\mu}_\psi &=& -\Gamma \left( \phi_{,\mu} \phi^{,\mu} \right) \left( \psi_{,\mu} \psi^{,\mu} \right) u^\mu_\psi \, . \label{eq:cov-coupling-bb}
\end{eqnarray}
\end{subequations}
Thus, the covariant transfer rates~\eqref{eq:cov-coupling-b} are again symmetrical 
with respect to the fields $\phi,\psi$.

\section{Unidirectional energy transfer between scalar fields}\label{App:unidirectional}
The energy coupling $Q = \beta \dot{\psi} \dot{\phi}$ is symmetric under the fields $\psi$ 
and $\phi$, which means that the transfer of energy proceeds as long as the fields roll 
down their potentials with a non-negligible kinetic energy. However, the transfer of
energy can be interrupted for one of the fields if there are rapid oscillations in their
evolution, as we shall show with a simple example.

The KG equations~\eqref{eq:Klein-Gordon} are written explicitly as,
\begin{subequations}
    \begin{eqnarray}
        \ddot{\psi} + 3H \dot{\psi} +\partial_{\psi} V(\psi) &=& - \beta \dot{\phi} \, , \\
        \ddot{\phi} + 3H \dot{\phi} +\partial_{\phi} V(\phi) &=& \beta \dot{\psi} \, .
    \end{eqnarray}
\end{subequations}
We assume the evolution of the fields at late times, so that the SFDM field rapidly
oscillates around the minimum of its potential. Under this circumstance, we can see
that on average $\langle \beta \dot{\phi} \rangle \simeq 0$, which means that there is no energy extraction from the EDE field and then it effectively evolves as an
uncoupled field. In contrast, for the SFDM field we find that $\beta \dot{\psi} \neq 0$, 
as long as $\dot{\psi} \neq 0$ for the EDE field.

More precisely, the new KG equations are,
\begin{subequations}
    \begin{eqnarray}
        \ddot{\psi} + 3H \dot{\psi} +\partial_{\psi} V(\psi) &=& 0 \, , \\
        \ddot{\phi} + 3H \dot{\phi} +\partial_{\phi} V(\phi) &=& \beta \dot{\psi} \, .
    \end{eqnarray}
\end{subequations}
Interestingly enough, we have the situation in which one of the fields evolves
freely, while at the same time the other field has its evolution modified by a non-homogeneous 
term in the KG equation for the field $\phi$.

Notice that this is consistent with the equations of motion in terms of the energy
densities of the fields. If we recall that 
\begin{subequations}
    \begin{eqnarray}
        \dot{\rho}_\psi &=& - 3H (\rho_\psi + p_\psi) - \beta \dot{\phi} \dot{\psi} \, , \\
        \dot{\rho}_\psi &=& - 3H (\rho_\phi + p_\phi) + \beta \dot{\phi} \dot{\psi} \, .
    \end{eqnarray}
\end{subequations}
Taking into account the rapid oscillations of the field $\phi$, we can see that $\langle \beta \dot{\phi} \dot{\psi} \rangle \simeq \beta \dot{\psi} \langle \dot{\phi} \rangle \simeq 0$, and
then at the level of the energy densities, the fields on average do not exchange energy.

\bibliography{references}        

\end{document}